\documentclass[prl,twocolumn,superscriptaddress,showpacs,floatfix]{revtex4}

\usepackage{graphicx}
\usepackage{amsmath}

\begin{document}

\title{Temperature dependent fluorescence in
disordered Frenkel chains: interplay of equilibration and local
band-edge level structure}

\author{M. Bednarz}
\affiliation{Institute for Theoretical Physics and  Material
Science Center, University of Groningen, Nijenborgh 4, 9747 AG
Groningen, The Netherlands}

\author{V.A. Malyshev}
\affiliation{S.I. Vavilov State Optical Institute'', Birzhevaya
Liniya 12, 199034 Saint-Petersburg, Russia}

\author{J. Knoester}
\affiliation{Institute for Theoretical Physics and  Material
Science Center, University of Groningen, Nijenborgh 4, 9747 AG
Groningen,
The Netherlands}

\date{\today}

\begin{abstract}

We model the optical dynamics in linear Frenkel exciton systems
governed by scattering on static disorder and lattice vibrations,
and calculate the temperature dependent fluorescence spectrum and
lifetime.  The fluorescence Stokes shift shows a nonmonotonic
behavior with temperature, which derives from the interplay of the
local band-edge level structure and thermal equilibration. The
model yields excellent fits to experiments performed on linear dye
aggregates.

\end{abstract}


\maketitle

Optical dynamics and emission of light in solid state systems are
topics of considerable fundamental and technological importance. A
variety of materials is currently studied as light emitting
devices. Examples are semiconductor quantum wells and wires
\cite{Gudiksen02}, conjugated polymers \cite{Gupta02}, and
molecular aggregates \cite{Spano00}. Closely related is the quest
for understanding the optical dynamics in photosynthetic antenna
systems \cite{Renger00,Trinkunas01}. The fundamental understanding
and description of the emission properties of all these systems
are challenging, because they involve a complicated interplay of
processes. The emission is usually caused by excitons and is
particularly sensitive to the scattering of these quasi-particles
on static disorder, which may lead to their localization, and
lattice vibrations. The latter results in vibronic relaxation and
hopping of the excitation between (localized) exciton states,
which gives rise to spatial and spectral diffusion.  These
diffusion processes compete with spontaneous emission and
nonradiative loss. Depending on the details of this competition,
the excitons may or may not reach thermal equilibrium before
emission occurs. Temperature plays an important role in this
competition. The thermal destruction of cooperative emission in
molecular aggregates \cite{deBoer90}, quantum wells
\cite{Feldmann87}, and quantum wires \cite{Citrin92,Oberli99} is a
well-known example of an interesting temperature dependent effect.
The Stokes shift of the exciton fluorescence may exhibit
interesting thermal behavior as well
\cite{Zimmerman97,Scheblykin01}.

In this Letter, we theoretically study the exciton dynamics and
the resulting emission properties of linear chains carrying
small-radius excitons. We account for both exciton localization by
static disorder and scattering of the excitons on lattice
vibrations.  The interplay of nonequilibrium conditions and the
local band-edge level structure of disordered chains leads to
interesting effects in the emission. A comparison to experiment is
made.

We consider a chain of $N$ molecules described by the Frenkel
exciton Hamiltonian
\begin{equation}
H = \sum_{n=1}^N \> \epsilon_n |n\rangle \langle n| +
\sum_{n,m}^N\> J_{nm} \> |n\rangle \langle m| \ , \label{Hex}
\end{equation}
where $\epsilon_n$ is the excitation energy of the $n$th molecule
and $|n \rangle$ denotes the state in which molecule $n$ is
excited and all others are in their respective ground states.  In
order to model the effect of random solvent shifts, the
$\epsilon_n$ are assumed to be mutually uncorrelated Gaussian
stochastic variables with mean $\epsilon_{0}$ and standard
deviation $\sigma$. The hopping integrals $J_{nm}$ are nonrandom
and originate from dipolar coupling: $J_{nm} = - J/|n-m|^{3}$,
$J_{nn} \equiv 0$. Here, $J$ is positive, implying that the
optically allowed exciton states reside at the bottom of the
exciton band.

We will assume that the coupling of the excitons to lattice
vibrations is weak. It is then appropriate to use a basis of
eigenstates of $H$. This basis consists of $N$ exciton states,
which are labeled by $\nu=1,\ldots,N$. The energy of the $\nu$th
exciton, $E_{\nu}$, is the $\nu$th eigenvalue of the $N \times N$
matrix $H_{nm} = \langle n| H|m \rangle$, while the $n$th
component of the corresponding eigenvector, $\varphi_{\nu n}$,
gives the amplitude of the $\nu$th exciton state on molecule $n$:
$|\nu\rangle = \sum_{n=1}^N \varphi_{\nu n}|n\rangle$. The
disorder $\sigma$ leads to localization of the eigenstates and
gives rise to a tail in the density of states (DOS) below the bare
band edge \cite{Schreiber82,Fidder91a}. This tail is of particular
interest for optical properties, as the states residing in it
carry most of the oscillator strength, $F_\nu=|\sum_n \varphi_{\nu
n}|^2$.  While the DOS accumulated over the entire chain is a
smooth function of energy, locally it exhibits structure
\cite{Malyshev9501}. This local level structure is of key
importance to the low-temperature dynamics and is described below.

\begin{figure}[ht]
\centering
\includegraphics[width=6cm]{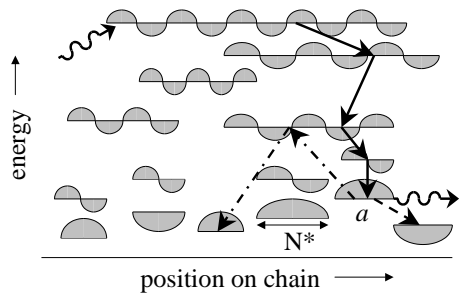}
\caption{{\protect\footnotesize} Schematic picture of the
localized exciton wave functions near the lower band edge in a
disordered linear chain. Also indicated are the optical pumping
and emission processes (wavy arrows) and various intraband
relaxation processes (dashed and dash-dotted arrows).}
\label{figstates}
\end{figure}

The exciton states in the tail of the DOS are localized on finite
segments, which near the band edge have a typical size $N^*$ (see
Fig.~\ref{figstates}). Within its own segment, each one of these
states is the lowest-energy eigenstate. They have wave functions
without nodes in the segments and carry an oscillator strength
$F_\nu \approx N^*$. Their spontaneous emission rates,
$\gamma_\nu=\gamma_0 F_\nu$ (with $\gamma_0$ the emission rate of
a single molecule), thus scale like $N^*$.  This is referred to as
superradiant enhancement \cite{deBoer90}. Bottom states on
neighboring segments have small overlap. Often a second,
higher-energy, exciton state may be distinguished on a segment,
separated from the bottom state by an energy of the order of
$3\pi^2 J/{N^*}^2$, as is dictated by energy quantization. These
states have one node within the segments and small oscillator
strengths. Still higher local states are usually delocalized over
two or more segments and carry negligible oscillator strengths.
The local level structure is washed out in the average over the
entire chain due to the fact that the distribution of energies of
the bottom segment states also has a width of the order of $3\pi^2
J/{N^*}^2$ \cite{Malyshev9501}.

Weak exciton-phonon coupling leads to scattering between the
exciton eigenstates. To describe this dynamics, we will use the
Pauli master equation for the populations, $P_\nu (t)$, of the
localized exciton states:
\begin{equation}
{\dot P}_\nu = R_\nu(t) -\gamma_\nu P_\nu + \sum_{\mu = 1}^N
(W_{\nu\mu}P_{\mu} - W_{\mu\nu}P_\nu) \, . \label{Pnu}
\end{equation}
Here, the dot denotes the time derivative, $R_\nu(t)$ is the
optical pumping rate, $\gamma_\nu$ is the above defined
spontaneous emission rate, and $W_{\nu\mu}$ is the transition rate
from the localized state $\mu$ to the state $\nu$ induced by the
exciton-phonon scattering.  For the latter, we will use
\begin{equation}
W_{\nu\mu} = W_0\ S(E_\nu - E_{\mu})\ G(E_\nu - E_{\mu})
\,\sum_{n=1}^N \varphi_{\nu n }^2 \varphi_{\mu n}^2 \, ,
\label{Wnumu}
\end{equation}
where the constant $W_0$ characterizes the scattering, the sum
over sites is the overlap integral of exciton probabilities for
the states $\mu$ and $\nu$, and $S(E_\nu - E_\mu)$ describes that
part of the $|E_\nu - E_\mu|$ dependence of $W_{\nu\mu}$ that
derives from the energy dependence of the exciton-phonon coupling
constant and the phonon density of states. Finally, $G(E) = n(E)$
if $E > 0$ and $G(E) = 1+n(-E)$ if $E < 0$, with $n(\Omega) =
[\exp(\Omega/k_{\mathrm {B}}T) - 1]^{-1}$ the mean occupation
number of a phonon state with the energy $\Omega$. Equation
(\ref{Wnumu}), which meets the principle of detailed balance, is
the generic result from first-order perturbation theory in the
exciton-phonon interaction \cite{Leegwater97,Shimizu01,Bednarz02}.

We have numerically simulated the above model of intra-band
exciton dynamics for scattering on acoustic phonons of the chains'
host medium. We have set $S=|E_\nu - E_\mu|/J$, which properly
accounts for a decrease of the coupling of excitons to acoustic
phonons at small wave vectors and prevents the divergence of
$W_{\nu\mu}$ at small energy differences.  We will illustrate the
salient results of our model by using parameters that are relevant
to molecular aggregates of the dye THIATS: $\epsilon_0 = 18138$
cm$^{-1}$, $J = 740$ cm$^{-1}$, and $\gamma_0 = (2/31) \times
10^8$ s$^{-1}$ \cite{Scheblykin01,Basko03}. These parameters are
fixed; the two free parameters are $\sigma$ and $W_0$, for which
we will use the values that yield the best fit to experimental
data: $\sigma = 0.23J$ and $W_0 = 30J$, respectively ({\it vide
infra}). A more elaborate analysis of our model, covering a wider
range of values for $\sigma$ and $W_0$, will be discussed
elsewhere \cite{Bednarz03}. The motivation to focus on the example
of THIATS aggregates, is that a large number of temperature
dependent spectroscopic data are known for these systems
\cite{Scheblykin00,Scheblykin01}.  We will limit our attention to
temperatures small compared to the Davydov splitting of 3000
cm$^{-1}$ that exists for these aggregates between the so-called J
band at the lower exciton band edge and the so-called H band at
the upper band edge \cite{Scheblykin01,Basko03}. For these
temperatures our effective model with one molecule per unit cell
may be used \cite{note}.  We have used a chain size of $N=500$ in
our simulations, which far exceeds the typical band-edge exciton
size $N^* = 30$ for $\sigma = 0.23J$.

We will first address the absorption and steady-state emission
spectra, which are calculated through
\begin{subequations}
\begin{equation}
A(E) = \frac{1}{N}\Big\langle\sum_{\nu=1}^{N} \gamma_{\nu}
\delta(E - E_{\nu}) \Big\rangle \ , \label{A}
\end{equation}
\begin{equation} I^{\mathrm{st}}(E) = \frac{1}{N}\Big
\langle\sum_\nu \gamma_\nu P_\nu^{\mathrm{st}} \delta(E -
E_\nu)\Big\rangle \ , \label{I}
\end{equation}
\end{subequations}
respectively. Here, the angular brackets denote the average over
the disorder realizations. Furthermore, $P_\nu^{\mathrm{st}}$ is
the steady-state solution of the master equation Eq.~(\ref{Pnu})
with the source term given by $R_\nu(t)= F_\nu$ if $E_\nu$ falls
inside a narrow window in the blue wing of the J band and
$R_\nu=0$ otherwise. This blue-tail excitation condition agrees
with common experimental conditions.

\begin{figure}[ht]
\centering
\includegraphics[width=7cm]{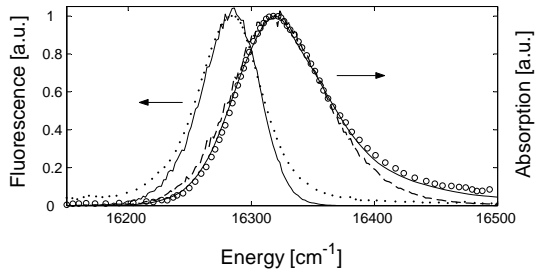}
\caption{{\protect\footnotesize} Calculated absorption and
fluorescence spectra (solid lines) at 10 K for the model
parameters described in the text, compared to the J band ($\circ$)
and fluorescence spectrum ($\bullet$) for THIATS aggregates
measured at $T=10$ K \cite{Scheblykin01}. The dashed line is the
fluorescence spectrum calculated for the same parameter set,
except that now $W_0=10^{-3}J$.} \label{figspectra}
\end{figure}

The spectra calculated for $T=10$~K are shown by solid lines in
Fig.~\ref{figspectra}, together with the J band and fluorescence
spectrum measured for THIATS aggregates at the same temperature
(dots) \cite{Scheblykin01}.  Clearly, both experimental spectra
are fitted very well within our model. The main deviations occur
in the tails, which are quite sensitive to the precise nature of
the disorder (diagonal versus off-diagonal disorder, no perfect
alignment of molecular dipoles). We note that the fluorescence
band is shifted to the red (Stokes shift) and narrowed with
respect to the J band. Keeping in mind that the absorption band is
dominated by the set of superradiant bottom states of the various
localization segments on the chain, this is an indication that
after the fast initial relaxation to a particular bottom state
(e.g., the level $a$ in Fig.~\ref{figstates}), the excitons have a
chance to relax to even lower states. This is a signature of
migration to other localization segments, a process which at low
temperatures is hindered by the small overlap between exciton wave
functions of different segments.  The fact that, in spite of this,
migration may occur before radiative emission takes place, is due
to the large value of $W_0$, which brings the system into a regime
that we refer to as the fast-intersegment-relaxation limit. To
illustrate the role of $W_0$, we also present in
Fig.~\ref{figspectra} (dashed line) the simulated fluorescence
spectrum for $W_0=10^{-3}J$ (all other parameters as above). In
this slow-relaxation limit the fluorescence spectrum closely
follows the absorption band and no Stokes shift occurs.

At zero temperature, the possibility for the excitation to migrate
to other localization segments, after the initial relaxation
process, is very small.  The reason is that the migration may then
only occur via a jump to a lower exciton state on a neighboring
segment, for which at least some (albeit small) wave function
overlap exists (cf. the process denoted by the dashed arrow in
Fig.~\ref{figstates}). However, the probability of finding a lower
exciton state on a neighboring segment is small, because the
initial state already lies in the tail of the DOS. Therefore, in
practice, low-temperature exciton migration is limited to at most
one jump. As a consequence of this blockade for diffusion, there
is not much opportunity for the excitation to explore the full DOS
during its lifetime and to equilibrate to the lowest state on the
disordered chain.  Thus, even for large $W_0$, the zero
temperature Stokes shift is limited to values of the order of the
J band width.

\begin{figure}[ht]
\centering
\includegraphics[width=7cm]{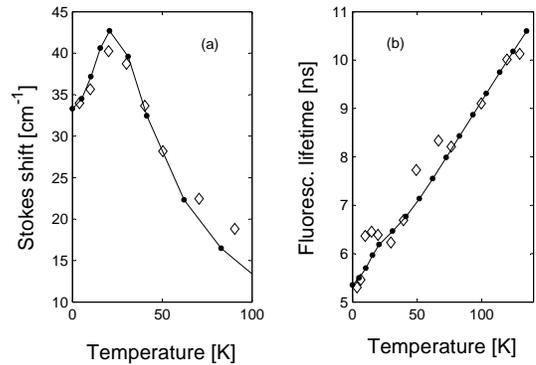}
\caption{{\protect\footnotesize} (a) Calculated Stokes shift of
the fluorescence spectrum as a function of temperature for the
parameter set described in the text ($\bullet$). The solid line is
a guide to the eye. Diamonds mark the measured Stokes shift for
THIATS aggregates reported in Ref.~\cite{Scheblykin01}. (b) As in
(a), for the fluoresecence lifetime instead of the Stokes shift.
Diamonds are the experimental data from Ref.~\cite{Scheblykin00}.}
\label{figStokestau}
\end{figure}

Upon increasing the temperature, higher states may thermally be
excited and one expects the Stokes shift to decrease and
eventually vanish. This is the generic behavior for single
molecules. Our simulations show that at elevated temperatures this
also occurs for the exciton system (Fig.~\ref{figStokestau}(a)).
However, our results also reveal a counter-intuitive intermediate
regime, where the Stokes shift increases with temperature. This
nonmonotonic behavior is intimately related to the interplay of
equilibration and the local level structure near the exciton band
edge. Upon slightly increasing the temperature from zero, a small
fraction of the population is thermally raised from the bottom
state of a segment to a higher exciton state that overlaps with
the same segment. As this higher state generally overlaps with
more segments, this provides a means of distributing the
population over two or more neighboring segments. The total
process is denoted by the dash-dotted arrows in
Fig.~\ref{figstates}. This overlap-assisted activated migration is
slow, but it allows the excitation to probe more segments for a
low-energy position and circumvents to some extent the blockade
for migration which occurs at $T=0$ K. Hence, the excitons get
closer to thermal equilibrium, which at low temperatures leads to
a larger Stokes shift.  Such anomalous low-temperature behavior
has also been predicted and observed for Wannier excitons
migrating between local potential minima in narrow quantum wells
\cite{Zimmerman97}.

As we also see in Fig.~\ref{figStokestau}(a), our numerical
results, obtained for the same parameter set as was used to fit
the low-temperature spectra, agree very well with the experimental
data for THIATS aggregates reported in Ref.~\cite{Scheblykin01}.
This gives strong evidence for the validity of our dynamic model
and the concept of overlap-assisted thermally-activated
intersegment hopping. Using the typical local level separation $3
\pi^2 J/{N^*}^2$ and the superradiant emission rate $N^*
\gamma_0$, it may be shown that the temperature interval over
which the Stokes shift behaves in an anomalous way (i.e.,
increases with $T$) scales roughly linearly with the absorption
line width and only weakly (inverse logarithmically) depends on
the ratio $W_0/\gamma_0$ \cite{Bednarz03}.

We finally turn to the lifetime of the total fluorescence
intensity after pulsed excitation.  For the generally
non-exponential intensity trace $I(t) = -\langle\sum_\nu {\dot
P}_\nu(t) \rangle$, we define the lifetime as its first moment,
i.e., the expectation value of the photon emission time:
$\tau=\int_0^{\infty} \langle \sum_\nu P_\nu(t) \rangle {\mathrm
d}t$ \cite{Bednarz02}. Here, $P_\nu(t)$ is the solution of
Eq.~(\ref{Pnu}) with $R_\nu(t)=cF_\nu \delta(t)$ if $E_\nu$ falls
in a narrow window in the blue tail of the absorption band and
$R_\nu(t)=0$ otherwise. The normalization constant $c$ is such
that directly following the excitation pulse the total population
on the chain, $ \sum_\nu P_\nu $, equals unity.

The generic behavior of the lifetime as a function of temperature
is well known.  At low $T$, the decay is dominated by the
superradiant bottom states of the localization segments, leading
to a fluorescence lifetime of the order of $1/N^*\gamma_0$.  With
increasing $T$, the exciton population is shared by the
superradiant states as well as an increasing number of
higher-lying dark states. This leads to an increase of the
fluorescence lifetime, as has been observed for various exciton
systems
\cite{Feldmann87,deBoer90,Citrin92,Oberli99,Kamalov96,Scheblykin00}.
For THIATS aggregates the lifetime shows a nearly linear
temperature dependence from 0 to 130 K \cite{Scheblykin00} (see
Fig.~\ref{figStokestau}(b), diamonds). For BIC aggregates a
similar linear increase was observed and, based on the flat nature
of the DOS for ideal two-dimensional systems, this was put forward
as evidence for a two-dimensional structure of these aggregates
\cite{Kamalov96}. It turns out, however, that the interplay of the
(local) DOS and equilibration in our one-dimensional model with
disorder may equally well explain the quasi-linear behavior. The
solid line in Fig.~\ref{figStokestau}(b) gives our simulated
results, obtained for the same parameters that were used to fit
the low-temperature spectra of THIATS.  We conclude that the
model, without any new parameters, yields an excellent fit to the
experimental data.  We also mention that at small $W_0$ (slow
relaxation), the linear temperature dependence may be preceded by
a low-temperature plateau \cite{Bednarz03}, which is the type of
behavior that is observed for aggregates of pseudo-isocyanine
\cite{deBoer90}.

In summary, we have analyzed the temperature dependence of the
fluorescence in one-dimensional disordered exciton systems and
obtained excellent fits to experimental data for aggregates of the
THIATS dye molecules.  As far as we are aware, this is the first
model calculation that yields such good agreement with a variety
of fluorescence experiments on linear dye aggregates. Previous
calculations focused only on the temperature dependent lifetime
and did not account for exciton localization by disorder
\cite{Spano90}. Our approach, based on the master equation with
perturbatively calculated rates $W_{\nu\mu}$, is consistent as
long as the rates for scattering between the optically dominant
states do not exceed their energy separation. For the parameters
used for THIATS we find that, in spite of the seemingly large
value of $W_0$, at low temperature this criterion is just obeyed,
due to the small overlap integral $\sum_{n=1}^N \varphi_{\nu n }^2
\varphi_{\mu n}^2 \sim 1/N^*$ for the exciton wave functions near
the band edge. With increasing temperature, the scattering rates
increase and around 100 K they lead to a visible contribution to
the absorption line width (homogeneous broadening)
\cite{Scheblykin01}. At such high temperatures the master equation
description breaks down and the full exciton density matrix must
be considered instead \cite{Renger00}.

V.A.M. acknowledges support from la Universidad Complutense at the
initial stage of this work.


\begin{thebibliography}{99}
\bibitem{note} This effective model requires multiplying the real
        molecular oscillator strength by the fraction of the total
        oscillator strength that is accumulated in the J band (i.e., =
        1/31 \cite{Scheblykin01}), as was done in our choice of $\gamma_0$.
\end{thebibliography}
\end{document}